\documentclass[a4paper]{article}

\usepackage{INTERSPEECH2022}
\usepackage{subfigure}
\usepackage{multicol,multirow}
\title{Improving Transformer-based Conversational ASR by \\Inter-Sentential Attention Mechanism}
\name{Kun Wei$^1$, Pengcheng Guo$^1$, Ning Jiang$^2$}
\address{
    $^1$Audio, Speech and Language Processing Group (ASLP@NPU), School of Computer Science,\\ Northwestern Polytechnical University, Xian, China
    \\
    $^2$Mashang Consumer Finance Co., Ltd.
    }
\email{ethanwei@mail.nwpu.edu.cn, pcguo@nwpu-aslp.org, ning.jiang02@msxf.com}

\begin{document}

\maketitle

\begin{abstract}
Transformer-based models have demonstrated their effectiveness in automatic speech recognition (ASR) tasks and even shown superior performance over the conventional hybrid framework. The main idea of Transformers is to capture the long-range global context within an utterance by self-attention layers. However, for scenarios like conversational speech, such utterance-level modeling will neglect contextual dependencies that span across utterances. In this paper, we propose to explicitly model the inter-sentential information in a Transformer based end-to-end architecture for conversational speech recognition. Specifically, for the encoder network, we capture the contexts of previous speech and incorporate such historic information into current input by a context-aware residual attention mechanism. For the decoder, the prediction of current utterance is also conditioned on the historic linguistic information through a conditional decoder framework. We show the effectiveness of our proposed method on several open-source dialogue corpora and the proposed method consistently improved the performance from the utterance-level Transformer-based ASR models.
\end{abstract}

\noindent\textbf{Index Terms}:
End-to-end speech recognition, Transformer, Long context, Conversational ASR

\section{Introduction}

Context information plays an important role in ASR, especially in scenes that require inter-sentential information such as conversation since semantically related words, or phrases often reoccur across sentences~\cite{kim2019cross}. Typically, traditional hybrid acoustic-language ASR models usually rely on rich language models to model contextual information~\cite{mikolov2010recurrent,mikolov2012context,mnih2007three, ji2015document,liu2017dialog, xiong2018session}. Meanwhile, there are also several researches adopting context information particularly in end-to-end ASR by adding additional context to the decoder or simply concatenate multiple consecutive utterances as the input of an end-to-end model~\cite{kim2018dialog, masumura2019large, masumura2021hierarchical}. 


    Transformer~\cite{vaswani2017attention}, as the most successful attention-based end-to-end model, has recently received more attention due to its superior performance on a wide range of tasks including ASR~\cite{wang2019learning, raganato2018analysis, dong2018speech, karita2019comparative, luo2021simplified, gulati2020conformer}.
    However, since the computational and memory cost of self-attention is quadratic w.r.t the input sequence length, Transformer is hard to process long sequences and mainly models independent utterances. 
    
    Several studies in natural language processing (NLP) have been explored to utilize the long contextual information for Transformer~\cite{dai2019transformer, rae2019compressive, beltagy2020longformer, zhou2020informer}. Inspired by above studies in the NLP task, some approaches were also proposed to incorporate contextual information across successive input sequences in Transformer-based ASR~\cite{masumura2021hierarchical,hori2020transformer}, but these methods do not solve the problem of the high computational and memory cost, or have high model complexity.
    In this study, we propose a novel Transformer-based architecture to explicitly model the inter-sentential information for conversational ASR. Inspired by~\cite{he2020realformer}, we include a residual attention module in the encoder, which accelerates the convergence speed and well models the long-range global dependencies within each input sequence. Besides, to further transfer the contextual information of previous sentences, we also propose a novel context-aware residual attention module, which transfers contextual information through attention scores. For the decoder part, we use an additional context module to learn more inter-sentential information. By using the methods above, we introduce inter-sentential contextual information in the popular Transformer ASR model. We demonstrate the superiority of our approach on two dialogue benchmarks (speech from two speakers) HKUST and Switchboard, a lecture benchmark TED-LIUM2, and a dialog dataset DATATANG-dialog, with obvious error rate reduction and neglectable increase of computational cost and model complexity.
    

\section{Transformer and Conformer}

    Transformer~\cite{vaswani2017attention} is an attention-based end-to-end model, which consists of multi-block stacked encoder and decoder. Each block can be characterized by a multi-headed attention (MHA) module, a position-wise feed-forward (FFN) module, layer normalization (LN) layers, and residual connections.

The MHA module calculates the score through the vectors values $(\textbf{V}^h)$ and keys $(\textbf{K}^h)$, and assigns values to the output embeddings queries $(\textbf{Q}^h)$~\cite{he2020realformer}:
    \begin{equation}
      \text{MHA}(\textbf{Q}^h,\textbf{K}^h,\textbf{V}^h) = \text{Concat} (head_{1},...,head_{m})\textbf{W}^{O} ,
    \end{equation}
where $head_{i}=\text{Attn}(\textbf{QW}^{Q}_{i},\textbf{KW}^{K}_{i},\textbf{VW}^{V}_{i})$. 
Here, $\textbf{Q}$, $\textbf{K}$ and $\textbf{V}$ are matrices with dimension $d_{k}$,$d_{k}$ and $d_{v}$. $\textbf{W}^{Q}_{i},\textbf{W}^{K}_{i}$ and $\textbf{W}^{V}_{i}$ are matrices that maps three vectors to the \emph{i}-th head in the multi-head attention space. $\textbf{W}^{O}$ is a linear layer to transform the output after the stitching.

In the attention module, we use the traditional scaled dot-product attention module~\cite{vaswani2017attention} to calculate the attention scores, as shown below:
    \begin{equation} \label{eq:dot_product}
      \text{Attn}(\textbf{Q},\textbf{K},\textbf{V})=
            \text{Softmax}(\frac{\textbf{Q}\textbf{K}^{T}}{\sqrt[]{d_{k}}})\textbf{V},
    \end{equation}
    where $\frac{\textbf{Q}^h\textbf{K}^h{T}}{\sqrt[]{d_{k}}}$ is a matrix representing the attention scores for each query and key.

  The FFN module is composed of two fully-connected layers with a ReLU activation in between, as follows:
    \begin{equation}
      \text{FFN}(x)= \text{ReLU} (x\textbf{W}_{1}+b_{1})\textbf{W}_{2}+b_{2},
    \end{equation}

    In addition to the two sub-layers in each encoder block, the decoder inserts a third sub-layer, which performs multi-headed attention over the source and target sequences. 

Recently, Gulati et al.~\cite{gulati2020conformer} combined Transformer and Convolutional Neural Networks (CNN) as Conformer. The Conformer encoder adds a convolution module between MHA and FFN in Transformer encoder blocks, simultaneously capturing local and global contextual information and leading to superior performance in ASR tasks.

\section{Proposed Method}
\subsection{Residual Multi-headed Attention}
Residual multi-headeded attention (ResMHA) closely follows the same Post LN strategy as in~\cite{vaswani2017attention}, which normalizes the output at the end of each MHA or FFN module~\cite{he2020realformer}. ResMHA connects MHA of adjacent layers through attention scores, as shown in Fig. 1 (a). Formally, it uses the attention score \emph{Prev} of the previous layer as a conditional input to calculating the attention score of the current layer’s MHA. In particular, \emph{Prev} is the attention score before the Softmax operation.:
\begin{equation}
    \begin{split}
        \text{ResMHA} & (\textbf{Q}^h,\textbf{K}^h,\textbf{V}^h, \emph{Prev}) =  \\
        & \text{Concat}(head_{1},...,head_{m})\textbf{W}^{O},
    \end{split}
\end{equation}
where $head_{i}=\text{ResAttn}(\textbf{QW}^{Q}_{i},\textbf{KW}^{K}_{i},\textbf{VW}^{V}_{i}, \emph{Prev}_{i})$. 
Like $head_{i}$, $\emph{Prev}_{i}$ is the slice of $\emph{Prev}$. Then these residual attention scores,  which corresponding to the MHA heads, will be added to the current attention calculation through the Residual attention (ResAttn) module:  
\begin{equation}
\begin{split}
  \text{ResAttn} &(\textbf{Q},\textbf{K},\textbf{V},\emph{Prev})=
      \text{Softmax}(\frac{\textbf{Q}\textbf{K}^{T}}{\sqrt[]{d_{k}}}+\emph{Prev}_i)\textbf{V}.
\end{split}
\end{equation}

We apply it to speech Transformer, accelerating the convergence speed of the model during training and improving the final speech recognition accuracy.

\subsection{Context-aware Multi-headeded Residual Attention}
\begin{figure}[htb]
  \centering
  \includegraphics[width=\linewidth]{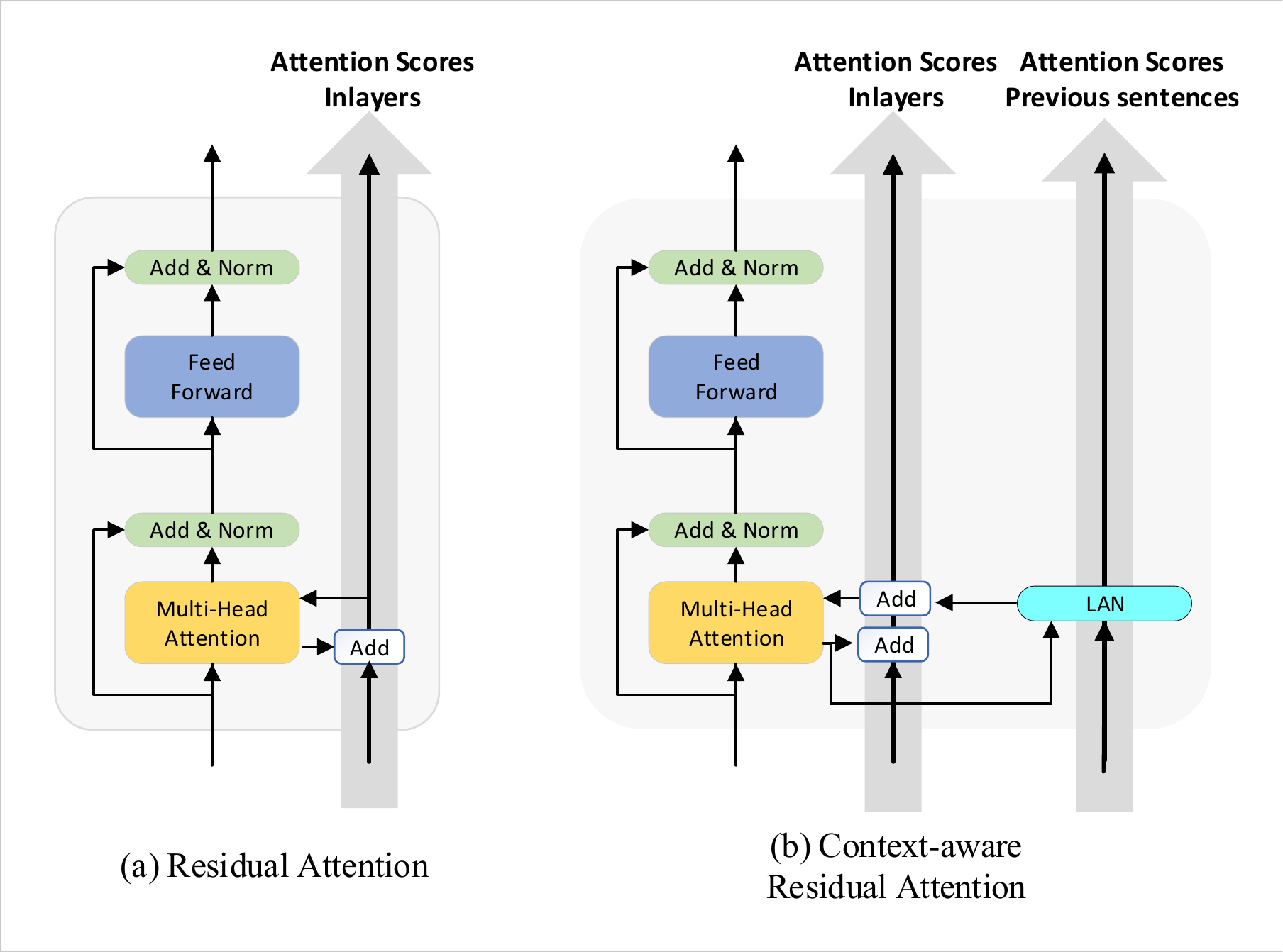}
  \caption{Details of the residual attention and context-aware residual attention.}
  \label{fig:realattention}
\end{figure}

Context-aware multi-headeded residual attention is designed on the basis of Residual attention~\cite{he2020realformer}, which adds a skip edge to connect multi-headed attention (MHA) modules adjacent layers, as shown in Fig. 1~(a). 
To capture contextual information across different consecutive utterances, an intuitive idea is to simply concatenate previous inputs with the current input to the encoder. Although, such method can give a slight performance improvement, it will also confront a large increment of memory cost and computation complexity. Thus, we propose a context-aware multi-headeded attention, which transfers the attention hidden states of the previous sentence in time order ($\emph{StateLS}$) to the current sentence and straightforwardly includes more contextual information during the training. Fig.~\ref{fig:realattention} (b) shows the details of our method. When we recognize the m-th sentence $X_{m}$ in the conversation and use the previous sentence $X_{m-1}$, or each $head_i$, the context-aware residual attention (CtxResAttn) can be formulated as:
\begin{equation}
\begin{split}
   \text{CtxResAttn}(\textbf{Q},\textbf{K},\textbf{V}, \emph{Prev}, \emph{PrevLS} (X_{m-1})) = \\
      \text{Softmax}(\frac{\textbf{Q}\textbf{K}^{T}}{\sqrt[]{d_{k}}}+ \emph{Prev}+\alpha \emph{PrevLS}(X_{m-1}))\textbf{V},
\end{split}
\end{equation}
where the CtxResAttn is our proposed context-aware residual attention module, $\emph{PrevLS}(X_{m-1})$ is the correlation attention score of previous one sentence before current input sentence $X_{m}$, and $\alpha$ is an interpolation factor to adjust the weight of historical information. The correlation attention score is calculated from pre-Softmax attention scores $X_{m-1}$ and $X_{m}$ as follows:
\begin{equation}
\begin{split}
     \emph{PrevLS}(X_{m-1})= \text{LAN}(s(X_{m-1}), s(X_{m})),
\end{split}
\end{equation}
where LAN is a linear layer, and $s(X_{m})$ is the pre-Softmax attention score of input $X_{t}$. Similar to the ResAttn, the new attention scores are applied on multi-head attention and passed over to the next layer.

\subsection{Conditional Decoder}

The decoder of Transformer contains a lot of linguistic information. In this part, we will describe our proposed conditional decoder, which adds the embedded vector to current input vector through an iterative attention block. As shown in Fig. 2, we recursively get the information of the previous vector, and then incorporate it with the current embedded input vector. Supposing we will look backward $n$ historic sentences when recognizing the $k$-th sentence in the conversation, the label texts $\mathbf{Y_{m-n}}$ and $\mathbf{Y_{m-n+1}}$ will be first processed by the Context-previous attention layer (CtxPrevAttn). 
\begin{equation}
\begin{split}\label{c1}
   &\text{CtxPrevAttn}(Y_{m-n}, Y_{m-n+1})= \\
    & \text{Attn}(\text{Embed}(Y_{m-n}), \text{Embed}(Y_{m-n+1}), \text{Embed}(Y_{m-n+1})),
\end{split}
\end{equation}
where the Embed is the word embedding layer and Attn is the dot-product attention mechanism, as described in Eq.~\ref{eq:dot_product}. Next, the output of the attention layer will be sent to a liner layer to get the position information:
\begin{equation}
\begin{split}\label{c2}
   & \text{Context}(Y_{m-n}, Y_{m-n+1})= \\
    & \text{LAN}(\text{CtxPrevAttn}(Y_{m-n}, Y_{m-n+1}), \text{Embed}(Y_{m-n+1})),
\end{split}
\end{equation}
where LAN is a liner transform layer. Then, we get the output of  
$\text{Context}(\text{Context}(Y_{m-n},Y_{m-n+1}),\text{Embed}(Y_{m-n+2}))$ in turn until the current $m$-th sentence. Finally, we send the contextual vector to the decoder of Transformer.
\begin{figure}[htbp]
  \centering
  \includegraphics[width=\linewidth]{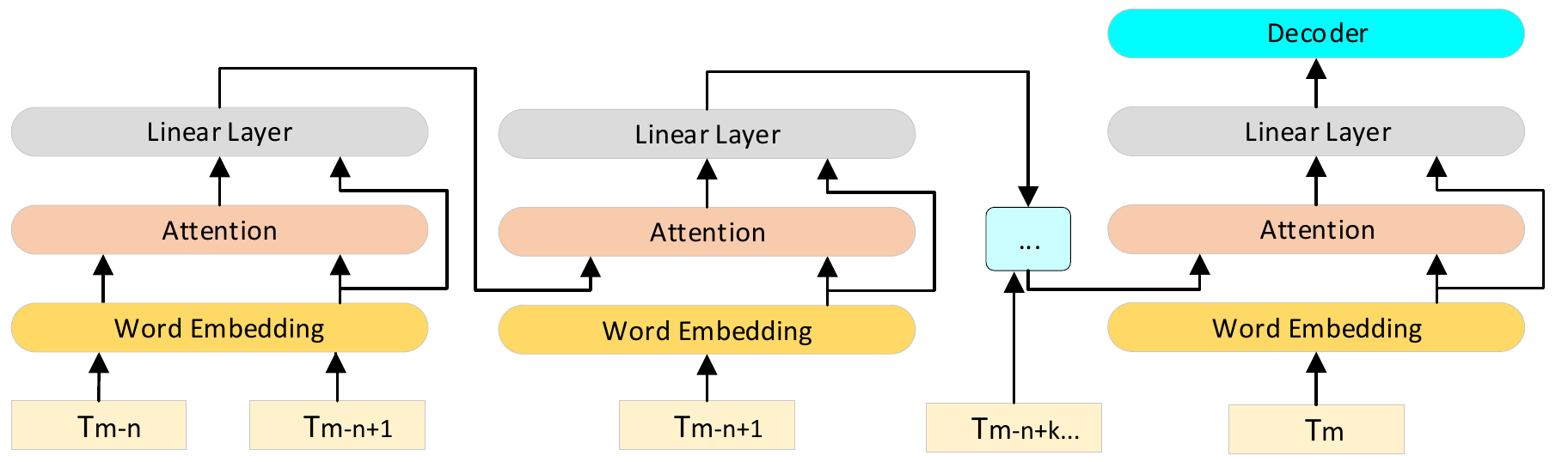}
  \caption{Conditional decoder with n previous sentences contextual information.}
  \label{fig:decoder}
\end{figure}

\section{Experiments}

\subsection{Dataset}

We conduct experiments on three dialogue datasets, including HKUST~\cite{liu2006hkust}, Switchboard~\cite{godfrey1992switchboard}, DATATANG-dialog, and a lecture dataset TED-LIUM2. Table 1 summarizes the information of each dataset. The 80-dimensional log-Mel filterbank(fbank) acoustic features plus 3-dimensional pitch features are used as the input features. 
For each corpus, the detail configurations of our Transformer model are same as ESPnet Transformer recipes~\cite{watanabe2018espnet} $(Enc =
12, Dec = 6, d^{ff} = 2048, H = 4, d^{att} = 256)$.
We use 1996 byte-pair-encoding (BPE) tokens~\cite{kudo2018sentencepiece} as output units for the English corpora Switchboard and TED-LIUM2, and characters for the Mandarin corpora HKUST and DATATANG-dialog.

\begin{table}[htbp]\centering
\caption{Dataset information.}
\begin{tabular}{l|l l l}
\hline
dataset         & lang & hours & test\_sets     \\ \hline
Switchboard     & en   & 260   & callhome/swbd  \\
TED-LIUM2          & en   & 213   & dev/test  \\
HKUST           & ch   & 200   & train\_dev/dev \\
DATATANG-dialog & ch   & 176   & train\_dev/dev \\ \hline
\end{tabular}
\end{table}

\begin{table*}[ht]\label{res}\centering
\caption{Results of proposed method. Numbers indicate WERs (\%) for SWITCHBOARD and TED-LIUM2, and CERs (\%) for HKUST and DATATANG(DATATANG-dialog) }
\begin{tabular}{c|c|c|c|c|c|c|c|c}
\hline
                     & \multicolumn{2}{c|}{Switchboard} & \multicolumn{2}{c|}{TED-LIUM2} & \multicolumn{2}{c|}{HKUST}    & \multicolumn{2}{c}{DATATANG} \\ \hline
 &
  \multicolumn{1}{c|}{callhome} &
  \multicolumn{1}{c|}{swbd} &
  \multicolumn{1}{c|}{dev} &
  \multicolumn{1}{c|}{test} &
  \multicolumn{1}{c|}{train\_dev} &
  \multicolumn{1}{c|}{dev} &
  \multicolumn{1}{c|}{train\_dev} &
  \multicolumn{1}{c}{dev} \\ \hline
Transformer baseline & 17.3            & 8.5            & 11.2           & 9.4           & 24.2          & 23.6          & 23.9          & 25.1          \\ \hline
+Proposed method     & \textbf{16.3}   & \textbf{8.3}   & \textbf{9.6}   & \textbf{8.7}  & \textbf{23.5} & \textbf{22.9} & \textbf{23.0} & \textbf{23.9} \\ \hline
Conformer baseline   &  15.6             & 8.4               & 10.2              & 9.0             & 20.8          & 20.0          & 17.4          & 18.1          \\ \hline
+Proposed method     & \textbf{15.1}     & \textbf{8.0}       & \textbf{9.7}             & \textbf{8.7}             & \textbf{19.9}          & \textbf{19.7}          & \textbf{17.0}          & \textbf{17.7}          \\ \hline
\end{tabular}
\end{table*}

\subsection{Experimental Setup}
We train the Transformer and Conformer~\cite{gulati2020conformer} models with the open-source end-to-end speech processing toolkit ESPnet~\cite{watanabe2018espnet, guo2020recent}. We use speed perturbation at ratio 0.9, 1.0, 1.1 for all corpora. 
We also apply SpecAugment~\cite{park2019specaugment} for additional data augmentation. Moreover, the Switchboard dataset is trained with more epochs than the other datasets. 
The baseline results are trained on independent sentence level, without speaker and context information.

We find that random input training, in other words, in the process of model training, the input sentences are not sent into the model in chronological order, will get better results than the time-ordered input. So instead of training with time-ordered input, we keep additional lists of dialog information during training. We send the features of previous sentences to the encoder to get the attention score. 
Since there is no need for backpropagation, the demand for computing resources is not significantly increased. 
For the decoder, 
we prepare two historical sentences, which means $n=2$,  for decoding every sentence.

In order to be consistent with the training input, in the decoding stage, we use the decode results of previous sentences to act as the previous text. For the first sentence in dialog, we just repeat this sentence to replace the position of the previous text. We use two previous sentences information in this paper when training models and the context-aware residual attention has a constant $\alpha$ value of 0.1.

When decoding, we average the best five models based on the validation loss for recognition on HKUST and DATATANG-dialog, while 10 models for Switchboard and TED-LIUM2. Besides, we use Long Short Term Memory Network language models (LSTM LM) to improve the recognition accuracy. The LM consist of four LSTM layers with 1024 units. The LM of HKUST, TED-LIUM2 and DATATANG-dialog are trained with the transcripts of training sets, and the Switchboard language model is trained with additional Fisher texts. ASR performance is measured by character error rate (CER) or word error rate (WER) depending on the language.

\subsection{Results and Analysis}
The results are shown in Table 2. For each dataset, the proposed method reduces ASR errors compared with the baseline, and the relative error rate reduction is up to 14\%. The DATATANG-dialog and TED-LIUM2 datasets get more error rate reduction. We suppose that those two datasets have greater topic coherence, which enhances the learning ability of the model to the corresponding keywords of specific domains. 

\subsection{Ablation Study}
We also conduct ablation experiments on HKUST and DATATANG-dialog datasets. In Table 3, $add\_1sen$ means using the proposed conditional decoder with $n=1$, real means using ResAttn and con means using CtcResAttn. Numbers indicate CERs (\%).
\begin{table}[ht]\centering
\caption{Ablation study on HKUST and DATATANG-dialog.}
\begin{tabular}{l|l|l|l|l}
\hline
             & \multicolumn{2}{c|}{HKUST} & \multicolumn{2}{c}{DATATANG} \\ \hline
             & train\_dev      & dev      & train\_dev        & dev       \\ \hline
Transformer Baseline     &  24.2               &    23.6      &     23.9              & 25.1          \\ \hline
+add\_1sen      &  24.0               &    23.5      &     23.7              & 25.2          \\ \hline
+add\_2sen      &  23.8               &    23.3      &     23.1              & 24.3          \\ \hline
+add\_2sen+real     & 23.6                &    23.3      &     23.1              & 24.1          \\ \hline
+add\_2sen+con & \textbf{23.5}                &    \textbf{22.9}      &     \textbf{23.0}              & \textbf{23.9}          \\ \hline
\end{tabular}
\end{table}
\subsubsection{Conditional Decoder}
 From the second row and the third row of Table 3, we can find that the attention decoder input with two previous sentences significantly improved the recognition accuracy. However, the attention decoder with one sentence gets negligible improvement and even the negative effect on the DATATANG-dialog dataset. We come to the conclusion that the longer contextual information, especially in the sentences with both sides of a conversation, can help the decoder to learn more language information.

 To verify the hypothesis, we performe speaker-dependent and speaker-independent experiments on dataset DATATANG-dialog. We add the previous sentence information of all speakers in the conversation in the speaker-independent training, and add the previous sentence information of current speaker in the speaker-dependent training. We verify the hypothesis in two different input methods, one is the time-order input of the dialogue, and the other one is the shuffle input. In the time-order training, we sort the input by speaker or simply sort by time to distinguish speaker-dependent or not. In the shuffle training, we keep a list of history context additionally, because we cannot get  historical information directly based on the shuffled input. We can see from Table 4 that, the results of speaker-dependent are worse than speaker-independent, both in model trained with time-ordered training dataset and shuffle training dataset. We can draw the conclusion that sentences that span both sides are better than the historical information on either side of the conversation.
 

\begin{table}[htbp]\centering
\caption{Speaker information study on DATATANG-dialog, SI is the speaker-independent training and SD is the speaker-dependent training, indicate CERs (\%).}
\begin{tabular}{c|l|c|c|c|c}
\hline
                          &    & \multicolumn{2}{c|}{time-order} & \multicolumn{2}{c}{shuffle} \\ \hline
                          &    & train\_dev         & dev        & train\_dev       & dev       \\ \hline
\multirow{2}{*}{baseline} & SI &  25.5                  &   26.8         & 23.9                 &  25.1         \\ \cline{2-6}
                          & SD &  25.9                  &  27.0          & 24.6                 &  25.7         \\ \hline
\multirow{2}{*}{proposed} & SI &  \textbf{23.7}  &  \textbf{24.8} &  \textbf{23.0} & \textbf{23.9} \\ \cline{2-6}
                          & SD &   24.2                 &  25.3          &  23.8        &  24.6         \\ \hline
\end{tabular}
\end{table}

\subsubsection{Context-aware Residual Attention}
The residual attention in encoder does not add any multiplication operations to the computational graph and improves the accuracy of the ASR task. Meanwhile, it achieves competitive results on ASR training with only 90\% of the number of epochs of the baseline. As shown in the forth row and the fifth row of Table 3, context-aware residual attention has greatly improved the recognition accuracy of our models. The context-aware residual attention adds previous context information and does not introduce too much redundant information into the encoder.

\subsection{The Impact of I-vector}
Previous work has shown that simply concatenating speaker related features, e.g. i-vector, with acoustic feature benefits conversational ASR ~\cite{audhkhasi2017direct,fan2019speaker}. I-vector can introduce additional speaker context and reduce the speaker mismatch between the training set and the test set. We further study our approach on the DATATANG-dialog dataset to see if there is still space for performance improvement when i-vector is adopted.

The i-vector estimator is trained with all the training data of DATATANG-dialog dataset. The training process follows the SRE08 recipe in Kaldi toolkit~\cite{povey2011kaldi}. A 2048 diagonal component universal background model (UBM) is first trained, and then 200-dimensional i-vectors are extracted and further compressed to 100 dimension by linear discriminant analysis (LDA) followed by length normalization. We concat the input fbank-pitch feature with 100-dimensional i-vector vector and send it to the network for training. It can be seen from Table 5 that the use of i-vector do lead to substantial performance gain, and our proposed method still can achieve extra performance improvement after adding i-vector as input.

\begin{table}[]\centering
\caption{CER (\%) results of i-vector and our proposed method on DATATANG.}
\label{tab:my-table}
\begin{tabular}{l|l|l}
\hline
                  & train\_dev    & dev           \\ \hline
Baseline              & 23.9          & 25.1          \\ \hline
    +proposed method          & 23.0          & 23.9          \\ \hline
    +i-vector     & 21.3          & 21.9          \\ \hline
    +i-vector+proposed method & \textbf{20.6} & \textbf{21.2} \\ \hline
\end{tabular}
\end{table}

\subsection{Attention Mechanism in Conditional Decoder}
We also analyze the capture ability improvement of recognition ability of keywords after adding the conditional decoder context module. We compare the attention scores of baseline decoder and conditional decoder in the third layers. We use the attention score of the first head in the decoder layer to plot the figure. Fig. 3 shows an example.

We can find that the dark color in Fig. 3(b) is concentrated on the diagonal while Fig. 3(a) has not yet formed a reasonable dissemination of attention. We can draw a conclusion that the conditional decoder with the context module can get accurate language information faster at a shallower level. Moreover, the dark blocks are better focused on the keywords mentioned in the previous conversation, which means the attention layer in Figure 3(b) improves the ability to perceive the keywords mentioned above.

\begin{figure}[htbp]
\centering
\subfigure[baseline]{\includegraphics[width=0.23\textwidth]{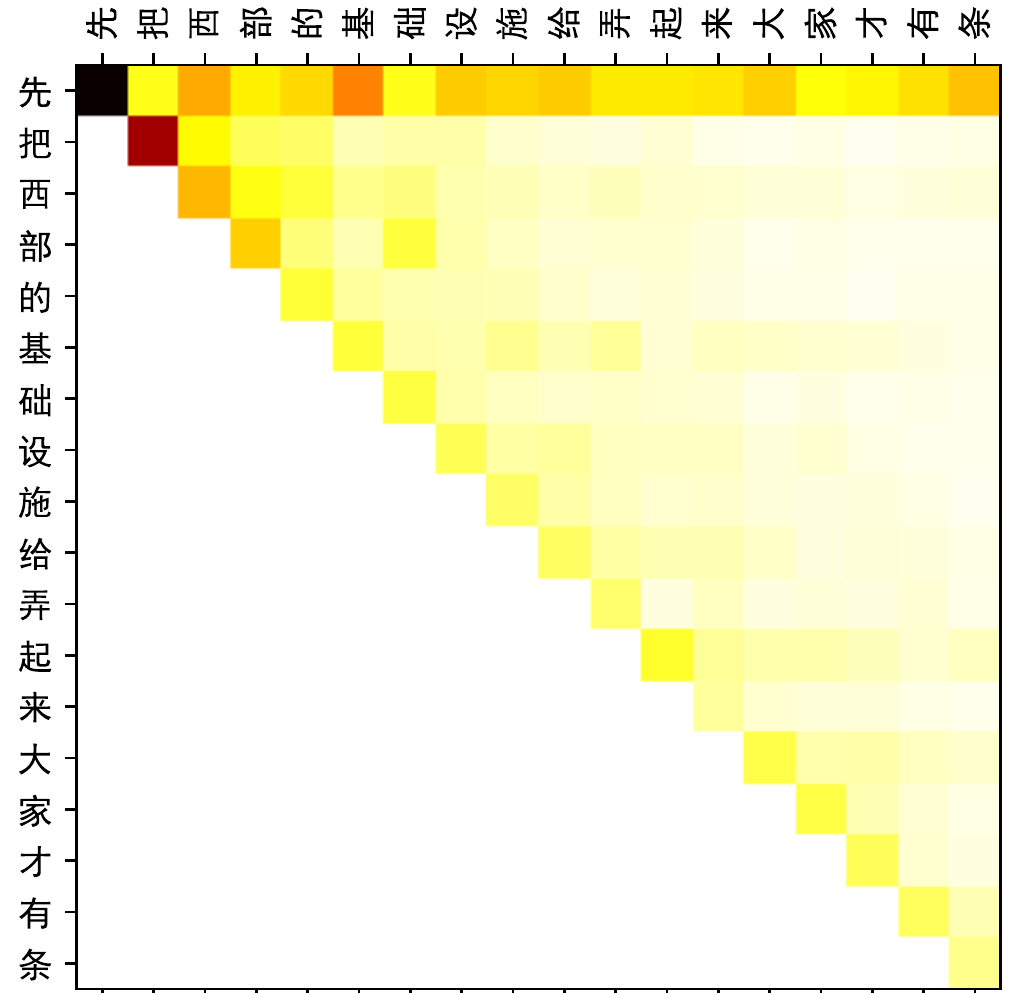}}
\subfigure[conditional decoder]{\includegraphics[width=0.23\textwidth]{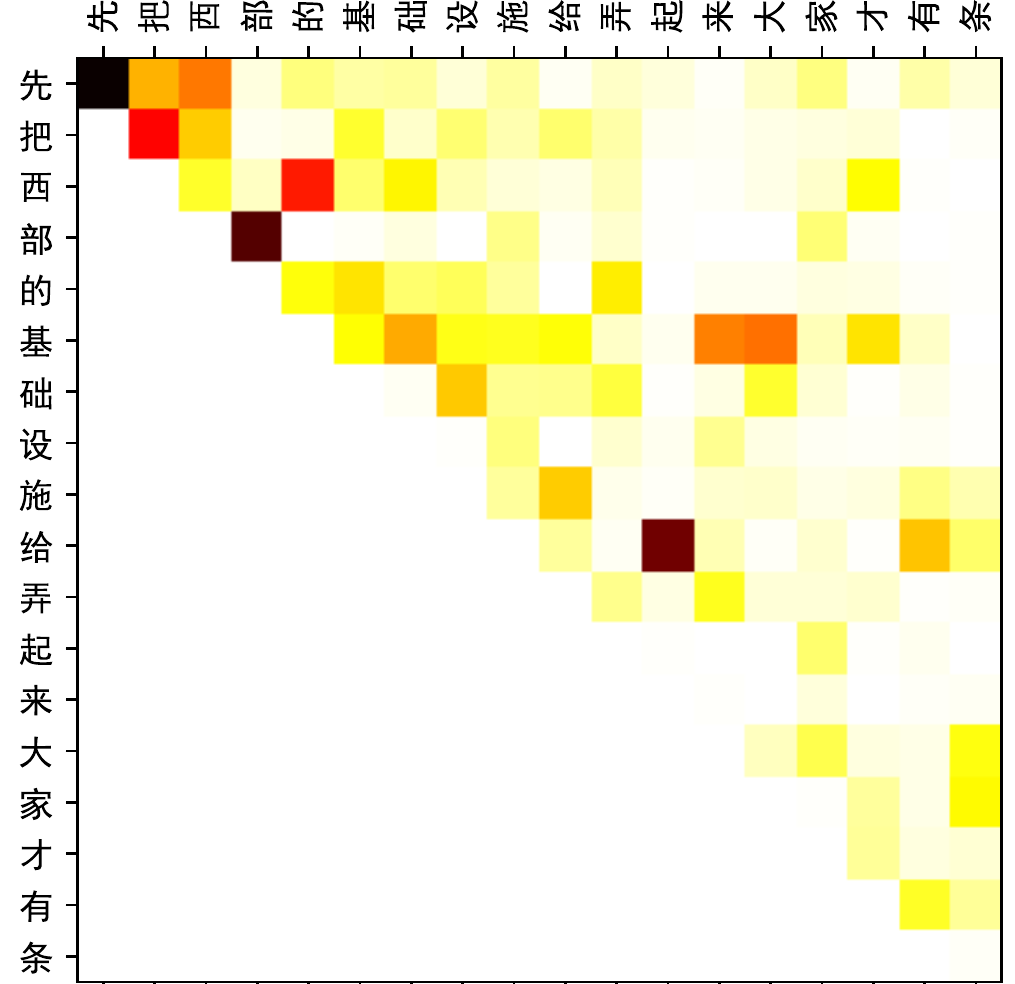}}
\caption{Attention scores of 3-th decoder layers, a darker color indicates a higher score for the character.}
\label{fig:1}
\end{figure}

\section{Conclusions}

In this paper, we design context-aware residual attention to get the contextual information without extra modules and parameters to the encoder. Moreover, the conditional decoder takes the text information from previous speech and improves the ability of the model to capture long contextual information. The experiments on four datasets demonstrated the effectiveness of our method in enhancing the prediction capacity in dialog ASR tasks. We will improve the decoding speed of our method and context-sensitive decoding strategies in our future work.

\bibliographystyle{IEEEtran}

\bibliography{mybib}


\end{document}